# The Use of Fuzzy Cognitive Maps in Analyzing and Implementation of ITIL Processes

**Hamid Zarrazvand [1], Mohammad Shojafar [2]**

[1] Computer Dept., Islamic Azad University, South Tehran Branch, Tehran, Iran,     E-mail: Hamidzar2002@gmail.com Tel: +98-912-622-2588 Fax: +98-112-523-2715

[2]  Computer Group, Young Researcher Community, Behshahr Branch, Iran,   E-mail: M.Shojafar@yahoo.com Tel: +98-935-846-8372

## Abstract

Information Technology Infrastructure Library (ITIL) is series of best practices that helps Information technology Organizations to provide Information technology (IT) services for their customers with better performances and quality. This article is looking for a way to implement ITIL in an organization and also using Fuzzy Cognitive Maps (FCM) to model the problem for better understanding of environment. ITIL helps to improve the performance of IT services in order to gain business objectives and Fuzzy Cognitive Maps will help to model the problem of needing ITIL processes for those objectives. First, it defines the concept of FCM and ITIL in two separate sections and then, it will describe the relationship and the way that FCM helps to implement ITIL. The paper will measure the cost of service support that is depended on the metrics like changes Authorization Degree, Process Oriented activities degree, Response time and Interrupt time.

This paper will be used as a part of gap analyzes step in implementing ITIL in each organizations.

## 1. Introduction

In these years the new challenge for IT companies is to manage the variety of services with an appropriate methodology or framework. Thus, to achieve this goal, several IT managers looking for a standard way to manage their services centrally and try to reduce the impact of IT Services interruptions in organizations.

There is a popular framework for managing IT services which is termed ITIL (IT Infrastructure Library) and was published by OGC (Organization Governance Commerce) in 1990. ITIL version 2 was released in 2001 and version 3 in summer of 2007 [1].

ITIL is the most popular framework for managing Information Technology services [2]. It defines standard processes for managing services, for example Incident Management is a process to manage the lifecycle of unwanted events in the environment.

As a matter of fact, today the biggest challenge for IT managers is to answer this question that how should they understand the environment and control all the elements with less costs and time? To gain this goal, implementing IT frameworks and trying to match business goals with IT Objectives is a common way.

This paper will define what FCM (Fuzzy Cognitive Mapping) can do to help IT Managers to design services base on their assets and Frameworks. It can provide a way to visualize the problem of managing services in IT Environment and let managers to solve this issue with a better understanding of their environment; Hence, it is going to assess the advantage of using FCM to implement ITIL framework in order to achieve business goals for organization, also, it will answer that how ITIL can help to find the best way of implementing ITIL base on organization objectives?

In order to achieve all the objectives which are discussed above, this paper is going to explain those in different sections. In Next Section, it defines ITIL disciplines and structure. FCM in real world described in Section 3. In Section 4, the paper approach the using of FCM/FRM to implement ITIL and the rules will be depicted. Then diagrams and models of proposed method based on "cost of service support parameters" will be illustrated. Finally, In Conclusion part, it described the goals and future view of ITIL clearly. Hence, in the next part ITIL disciplines will be explained.

## 2. Information Technology Infrastructure Library

ITIL is a set of concepts and practices for Information Technology Services Management (ITSM), IT development and IT operations.

ITIL gives detail descriptions for IT processes and provides comprehensive checklists, tasks and procedures that any IT organization can adjust to its requirements. ITIL was published as series of books; each covers an IT management topic. The names ITIL and IT Infrastructure Library are registered trademarks of the United Kingdom's Office of Government Commerce (OGC). It consists of publications giving guidance on the provision of Quality IT Services, and on the Processes and facilities needed to support them. ITIL has three foundation concepts:

1) *Service:* A Service provided to one or more Customers by an IT Service Provider.

2) *Service Provider:* A Service Provider that provides IT Services to Internal Customers or External Customers.

3) *Costumer:* Costumer is a one who consumes the service.

ITIL version 3 has five core disciplines and more than 15 disciplines which are included in core discipline. Below is the list of disciplines:

1. ITIL Service Strategy
    a. Definition of Business Service Requirement
    b. Determination of Market Space, IT Policies & Strategies
    c. Specify of Service Portfolio
    d. Demand Management
    e. Financial Management
2. Service Design [6]
    a. Service Level Management
        i. Service Catalog Management
        ii. Supplier Management
    b. Availability Management
    c. ITSCM
    d. Capacity Management
    e. Information Security Management
3. Service Transition
    a. Change Management
    b. Service Asset and Configuration Management
    c. Knowledge Management
    d. Release and Deployment Management
4. Service Operation
    a. Functions
        i. Service Desk
        ii. Operation Management
        iii. Technical Management
        iv. Application Management Lifecycle
    b. Processes
        i. Event Management
        ii. Incident Management
        iii. Request Fulfillment
        iv. Problem Management
        v. Access Management
5. Service Continual improvement
    a. 7-Step Improvement Process
        i. Define what you should measure
        ii. Define what you can measure
        iii. Gathering the data
        iv. Processing the data
        v. Analyzing the data
        vi. Presenting & using the information
        vii. Implementing corrective action

b. Deming Cycle and CSI Model

Figure 1 illustrates a view of ITIL Version 3 Process Diagram.

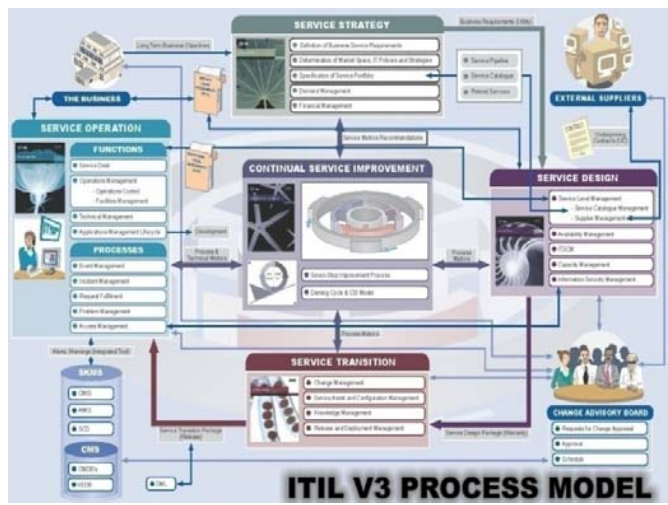

**Figure 1: ITIL V 3 Process Model [8]**

## 3. Fuzzy Cognitive Mapping

This section is going to define the concepts of FCM and its values. FCM visualize the problem and clarify it to understand the whole problem with diagrams of graphs. As a matter of fact, this concept looks for a directed graph to show the problem when it becomes hard to understand in usual methods. Here, there is a very important definition:

*"An FCM is a directed graph with concepts like policies, events and etc. as nodes and causalities as edges. It represents causal relationship between concepts "[3, 4]*

Let's take a detailed look at the concept, In FCM these definitions existed:

- When the nodes of the FCM are fuzzy sets then they are called as *fuzzy nodes*
- FCMs with edge weights or causalities from the set *{–1, 0, 1]* are called simple FCMs
- The matrix $E$ be defined by $E = (e_{ij})$ where $e_{ij}$ is the weight of the directed edge $E$ is called the adjacency matrix of the FCM, also known as the connection matrix of the FCM
- An FCM with cycles is said to have a feedback
- When there is a feedback in an FCM, i.e., when the causal relations flow through a cycle in a revolutionary way, the FCM is called a *dynamical system*

- Finite number of FCMs can be combined together to produce the joint effect of all the FCMs

Moreover, this section clarifies the properties and FRM.

### 3.1 Properties and models

FCM is a collection of classes which is drawn by circle or oval and casual relations between them. The directed edge $e_{ij}$ from causal concept $C_i$ to concept $C_j$ measures how much $C_i$ causes $C_j$. FCMs are used to model many varieties of problems like political, economic, social, organizational management and etc.

The edges $e_{ij}$ take values in the fuzzy causal interval $[-1, 1]$. $e_{ij} = 0$ indicates no causality, $e_{ij} > 0$ indicates causal increase $C_j$ increases as $C_i$ increases (or $C_j$ decreases as $C_i$ decreases). $e_{ij} < 0$ indicates causal decrease or negative causality. $C_j$ decreases as $C_i$ increases (and or $C_j$ increases as $C_i$ decreases). Simple FCMs have edge values in $[-1, 0, 1]$.

In FMC, it is possible to pass a state vector in order to find a second node, which will be resulted of the first node that passed the vector. For example on Figure 2 if it passes vector $(1, 0, 0, 0, 0)$ from node $c_1$ then, it will create a cycle between $c_1$ and $c_5$ which means increasing in population will increase unemployment $c_5$ is termed as a fixed point. Hence, you can pass a vector for each node to find the fixed points.

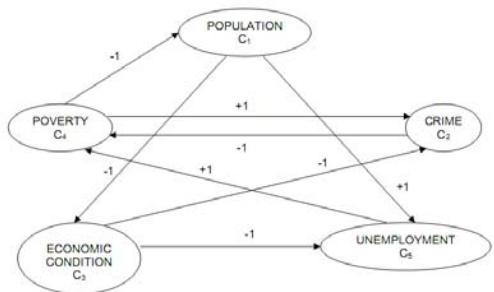

**Figure 2: FCM Model -1**

In the Figure 2, there is an example of FCM model which represent A Socio-economic model constructed with Population, Crime, Economic condition, Poverty and Unemployment as nodes or concept [11].

There are different types of expert and aspects to create the FCM, therefore, in this situation definition of different FCMs and combined them as one will be done in Eq. (1).

$$F = \sum F_i$$

Also, it can uses fuzzy sets for the causes that results in destination nodes, for example increasing the population will increases unemployment with $0.8$ rate and etc.

Here, another example of FCM to model and analyze business performance assessment which is so close to the subject of this paper. It is illustrated in Figure 3 [5].

For this kind of problems, a model was proposed by *Kardaras* and *Mentzas* which is called *Impact Analysis Model* (IAM). A part of IAM is to analyze the Information System or IT affect on business.

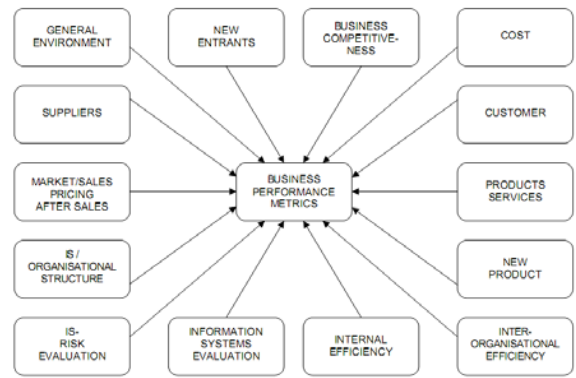

**Figure 3: FCM Model-2**

Figure 4 is an example of effects that have impact on IT opportunities and threats to change the performance of business.

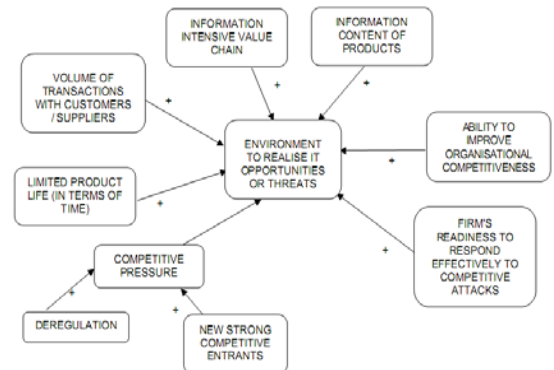

**Figure 4: FCM Model-3**

*3.2 FRM*

Here, it will show a brief view of problem in FRM; hence, let's take a quick look on this concept. FRM let us to make two disjoint groups of all the nodes (for example group *A* and *B* with domain space and a range space which are disjoint in the sense of concepts) and then it can result group B from A, in order to find the answer if some nodes of *A* happens which situation will resulted in group B.

There is an example in Figure 5. The Table 1 illustrated the Groups A and B. Group A is $D_i$ that considers teaching states and Group B $R_i$ stands for Rules.

**Table 1: Group A results Group B**

| Group A($D_i$) | | Group B($R_i$) | |
|---|---|---|---|
| $D_1$ | Teaching is good | $R_1$ | Good Student |
| $D_2$ | Teaching is poor | $R_2$ | Bad Student |
| $D_3$ | Teaching is mediocre | | |
| $D_4$ | Teacher is kind | $R_3$ | Average Student |
| $D_5$ | Teacher is harsh [or rude] | | |

This paper will use FRM to create a relationship between ITIL processes and inputs of FCM system (Fuzzy black box inputs) to show how implementing those processes will affect the results of FCM. It does it in this way because ITIL processes can not affect directly on output and FCM's output has been affected by inputs that related to those processes.

*3.3 FCM vs. FRM*

FCM is a graph (directed graph) that consists of two groups first one is concepts group (Nodes) and the second one is causalities (Edges). FCM uses edges to show the relationship between concepts and those concepts are something like policies, events and etc.

FRM is a technique to simply show how some works can result some effects (variables are in a fuzzy sets). It usually uses two groups to draw a diagram or table in order to show how the first group can cause the second group. FRM is not a graph like FCM; it consists of two sets that relate together. Indeed, this paper used FRM to demonstrate how implementing ITIL Processes can affect on the FCM concepts (organization policies) and then FCM will be used to show how policies can affect on the main goal/strategy of implementing ITIL.

Usually, when all concepts are in the same set, the FCM model is going to use, but, if there are concepts in different sets, it is better to use FRM

## 4. Implementing ITIL Using FCM/FRM (Approach)

First, it is required to know some templates of ITIL and the approaches of them; this will help to understand the relation of disciplines better, and then categorize ITIL processes into groups [9].

In the next figure it will show some reasons to introduce the approaches (it means the reasons cause to choose an approach). Hence, let take a quick look at an example which shows the relations between them that is illustrated in Table 2.

**Table 2: Relationships between reasons and approaches**

| Reasons | Approaches |
|---|---|
| Cost | Bare necessities |
| No Customer Support | Organic growth |
| ISO 2000 Limitations | Service support |
| Time Constraints | Service delivery |
| Ownership | Old version |
| Running out of steam | Old version + |
| Too Complex | Lifecycle |
| Old versions already implemented | Continual service improvement |
| | Service operation |
| | Service ownership |
| | Best practices |
| | Create template |

In this example the reason for implementing service support is costs (cost is the main reason of implementing ITIL service support for an organization).

*4.1 Service Support approach*

The main target of ITIL implementation in this paper is to use the disciplines for supporting IT services. It try to use those discipline that is related to support and it is not for those which are related to service design or service strategy etc; thus, the service support related processes will be picked up from a methodology that is explained in "ITIL lite : a road map to full or partial ITIL implementation"

For implementing ITIL as a service support approach it is needed to implement some essential processes of ITIL. These processes are listed below:

1. Incident Management[10]
2. Problem Management
3. Change Management[7]
4. Service Asset and Configuration Management
5. Service Desk (Function)

The processes which are known as service support process group are the most famous ones that are implemented in most IT organizations in a way of approaching ITIL. Hence, let's take a look at the relationships between them in Figure 5.

Figure 5 shows how Incident management interacts with all other processes and the fact that having a powerful team of service desk and Incident management help organizations to get to their approach as soon as it possible. After Incident management Service Asset and Configuration management is the next important process that helps organization for their approaches. Change management and Problem management are needed to complete the chain of diagnosing the interrupted services, which they were started from Incident management workflow. Finally, service desk is a core site of handling incidents which are created by end users, and it works as a function for IT department.

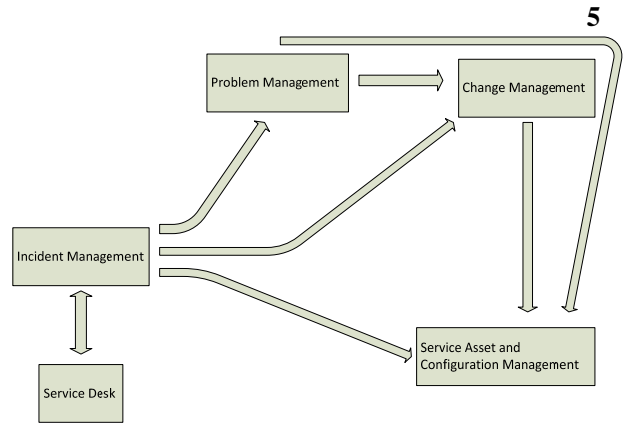

**Figure 5: Example of relations for ITIL**

*4.2 FCM*

Now, let's define how the using of FCM helps to understand the affect of each process on the way of implementing service support approach,

But before explanation, it is required to define the potential causes of using service support approach:

1. Decrease the cost of implementation
2. Quick response to end user
3. Fix interrupt as soon as it possible
4. Changes done by authorization
5. Process oriented system
6. Everything and each activities recorded

Hence, let see what happens in a simple FCM diagram for those causes in Figure 6. Figure 6 shows the relationships between the nodes that contain the causes and goals of implementing, for example, increasing/decreasing in response time will results in increasing/decreasing the cost of support and also increasing/decreasing the numbers of recorded events will decrease/increase the response time and so on.

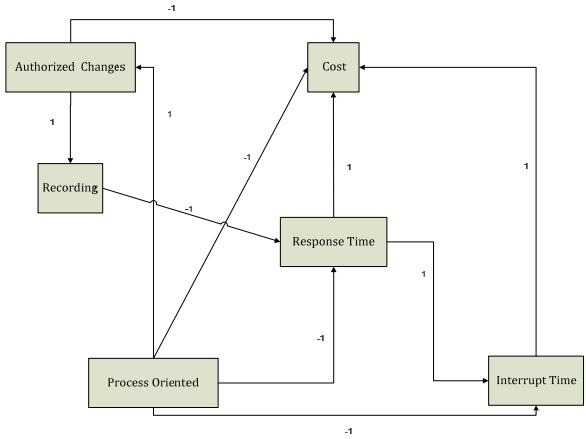

**Figure 6: ITIL Service Support approach FCM Model -1**

Here, is the adjacent matrix of goals FCM on Table 3.

**Table 3: Adjacent Matrix**

|  | Response Time | Cost | Interrupt | Process Oriented | Recording | Authorization |
|---|---|---|---|---|---|---|
| Response Time | 0 | 1 | 1 | 0 | 0 | 0 |
| Cost | 0 | 0 | 0 | 0 | 0 | 0 |
| Interrupt | 0 | 1 | 0 | 0 | 0 | 0 |
| Process Oriented | -1 | -1 | -1 | 0 | 0 | 1 |
| Recording | -1 | 0 | 0 | 0 | 0 | 0 |
| Authorization | 0 | -1 | 0 | 0 | 1 | 0 |

For the adjacent matrix, passing a vector (1, 0, 0, 0, 0, 0) to find a fixed point and a circle in Eq. (2).

$$C_1 E = (0,1,1,0,0,0) \quad \rightarrow \quad (1,1,1,0,0,0) = C_2$$
$$C_2 E = (0,2,1,0,0,0) \quad \rightarrow \quad (1,1,1,0,0,0) = C_3 = C_2$$

$$(2)$$

It means increase in response time will result increase in cost. And in another example if it passes $C_1$ as vector $(0, 0, 0, 1, 0, 0)$, for process oriented, the result is Eq. (3).

$$C_1 E = (-1,-1,-1,0,0,1) \quad \rightarrow \quad (0,0,0,1,0,1) = C_2$$
$$C_2 E = (-2,-2,-1,1,1,1) \rightarrow (0,0,0,1,1,1) = C_4 = C_3$$

$$(3)$$

When passing $C_1$ as vector $(0, 0, 1, 0, 0, 0)$ for Interrupts the result is Eq. (4).

$$C_1 E = (0,1,0,0,0,0) \quad \rightarrow \quad (0,1,1,0,0,0) = C_2$$
$$C_2 E = (0,1,0,0,0,0) \rightarrow \quad (0,1,1,0,0,0) = C_3 = C_2$$

$$(4)$$

After illustrating the relationships let redraw the diagram with the fuzzy sets relationships that are knowledge based relations and is shown here: (It uses the "too little ="0.1", little ="".03", usual ="".05", much="0.7", very much="0.9"" for the rate of memberships in set) in Figure 7.

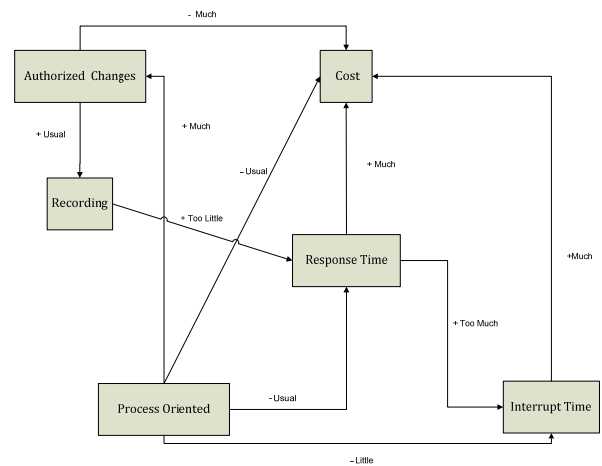

**Figure 7: ITIL Service Support approach FCM Model -2**

And, after replacement of "the rate of memberships" in the adjacent matrix table 3 will be converted to table 4.

$$(2)$$

**Table 4: Adjustment Matrix-Diffusification**

| | Response Time | Cost | Interrupt | Process Oriented | Recording | Authorization |
|---|---|---|---|---|---|---|
| Response Time | 0 | 0.7 | 0.9 | 0 | 0 | 0 |
| Cost | 0 | 0 | 0 | 0 | 0 | 0 |
| Interrupt | 0 | 0.7 | 0 | 0 | 0 | 0 |
| Process Oriented | -0.5 | -0.5 | -0.3 | 0 | 0 | 0.7 |
| Recording | -0.1 | 0 | 0 | 0 | 0 | 0 |
| Authorization | 0 | -0.7 | 0 | 0 | 0.5 | 0 |

Finally, it is better to find an optimum FCM diagram. To catching that goal it is required to keep important nodes and clear other nodes (by their rate), so choosing cost as a main output and "authorized changes", "interrupt time", "response time", "process oriented" as main input create another view of diagram. This optimum model is designed base on best practices and experiences. The following clearly shows it in Figure 8.

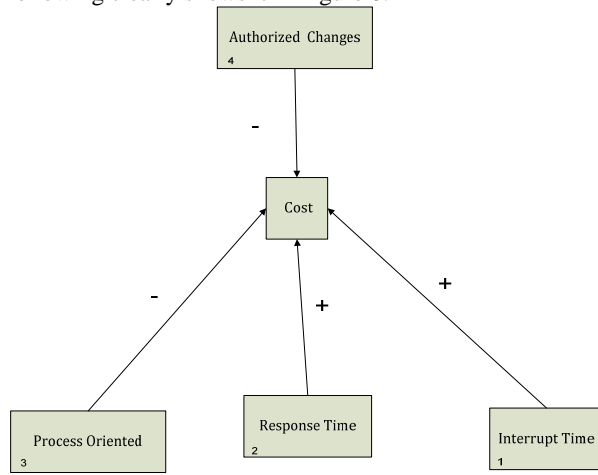

**Figure 8: ITIL Service Support approach FCM Model -3 (Optimum model)**

Also using fuzzy black box engine, to calculate the rules, can make some fuzzy decision box rules. Therefore, based on the proposed method this paper will be illustrated that problem in the next part as performance evaluation.

### 4.3 Performance Evaluation (Fuzzy black box)

The execution code of service support approach on ITIL was tested on PC with Notebook Core 2Duo 1.7GHZ, 2.0 Giga Byte Ram with Windows XP OS and programmed with Netbeans 6.8 (Java 1.6) Software and the diagrams are achieved on Fuzzy-tech 5.5 professional Edition Software.

There are some experimental rules that can help to understand the core of fuzzy black box. These are elicited from C1 (Interrupt Time), C2 (Response Time), C3 (Process Orientation Rate) and C4 (Authorized Change Rate) that mentioned before, showed in Table 5.

**Table 5: Rules Considered for Test based on C1, C2, C3, C4**

| |
|---|
| IF C1 is **little** and C2 is **little** and C3 is **normal** and C4 is **much** Cost will be **little** |
| IF C1 is **much** and C2 is **much** and C3 is **little** and C4 is **little** Cost will be **much** |
| IF C1 is **normal** and C2 is **normal** and C3 is **much** and C4 is **normal** Cost will be **normal** |
| IF C1 is **normal** and C2 is **normal** and C3 is **little** and C4 is **normal** Cost will be **normal** |
| IF C1 is **too little** and C2 is **too little** and C3 is **much** and C4 is **too much** Cost will be **too little** |
| IF C1 is **too little** and C2 is **too little** and C3 is **too much** and C4 is **too much** Cost will be **too little** |
| IF C1 is **too much** and C2 is **too much** and C3 is **little** and C4 is **too little** Cost will be **too much** |
| IF C1 is **too much** and C2 is **too much** and C3 is **too little** and C4 is **too little** Cost will be **too much** |

There are more rules that can imply the quantity of cost in fuzzy logic but the important issue is how these metrics in fuzzy can be converted to the mathematics metric. To solve this, it is possible to define a rate for each node, so, for example defining a budget or cost rate for cost of service support makes it clear that if the entire budget is used 100% of predicted budget is finished and so on, that map can be done for other 4 nodes (those will be defined later). Now, let's define a diagram to find how the percentage with the result of rules can be illustrated on Figure 9. 100 units tested in this experiment. Each color illustrated the terms for range of cost's result ("too little, little, normal, much, too much").

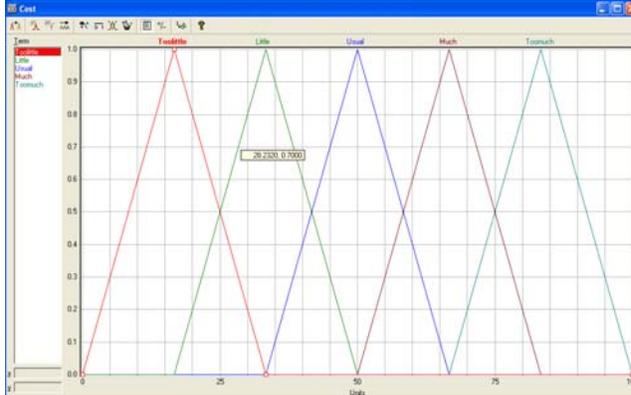

**Figure 9: Result of rules**

In this example, if the result be in "little" set with 0.7 degree of membership rate for cost, then the cost usage will be 28.23% of entire budget.

In this paper a sample code is attached for this example that had been described above, hence the program is coded in java and you can input the values in the following range (Table 6) and find the proper output for cost. In this sample only the above rules are defined (base on experiences had shown on **Table 5**), but there are more that can be included and make the fuzzy box engine more accurate.

**Table 6: Considered Parameters by Experiment**

| Input | (Rate) Range of values |
|---|---|
| Authorization Changes | 0% to 100% of changes |
| Interrupt Time | 0 to 1440 minutes per day |
| Process Orientation | 0% to 100% of works done in process oriented way |
| Response Time | 0 to 1440 minutes per day |
| Cost | 0% to 100% of Budget |

Table 6 defines the ranges of values of inputs and output for our fuzzy box.

Figure 10, 11, 12, 13 shows the fuzzification diagrams of input data and how it converts those values to fuzzy logic range.

Figure 10 illustrates the change authorization rate diagram that shows how the percentage of authorized changed in organization maps to fuzzy sets like "too little" or "much". Horizontal vector is started from 0 percentages to 100 percentages and vertical vector is started from 0 to 1 that shows the existence ratio of the point in each set

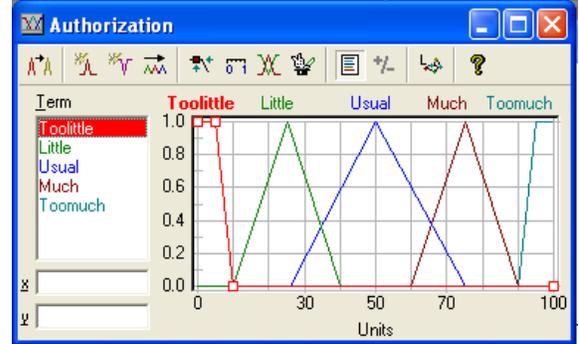

**Figure 10: Change Authorization (Authorized Changes)**

Figure 11 illustrates the service interrupt time rate that shows how the number of minutes of interrupted in organization's services per day maps to fuzzy sets like "too little" or "much". Horizontal vector is started from 0 minutes to 1440 minutes (1440 minutes is equal 1 day= 60*24) and vertical vector is started from 0 to 1 that shows the existence ratio of the point in each set.

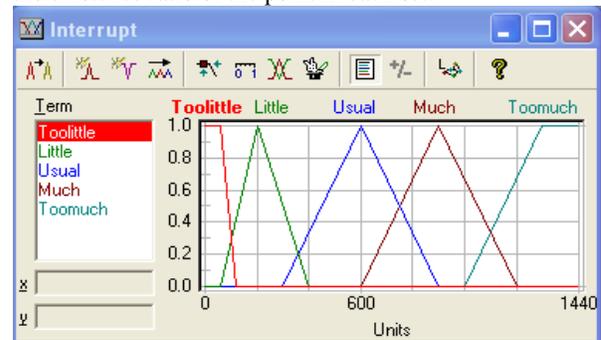

**Figure 11: Interrupt Time Rate**

Figure 12 illustrates the response time rate that shows how the number of minutes of response time in organization's services after an interrupt per day maps to fuzzy sets like "too little" or "much". Horizontal vector is started from 0 minutes to 1440 minutes (1440 minutes is equal 1 day= 60*24) and vertical vector is started from 0 to 1 that shows the existence ratio of the point in each set.

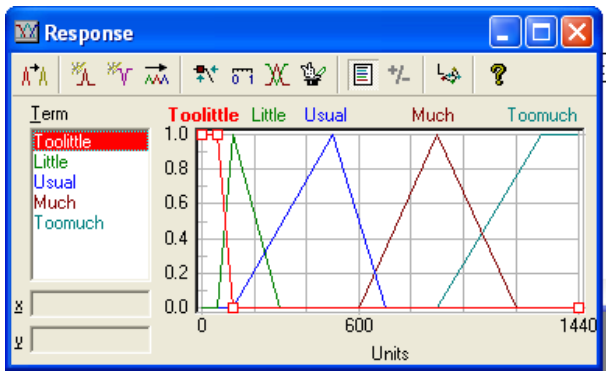

**Figure 12: Response Time Rate**

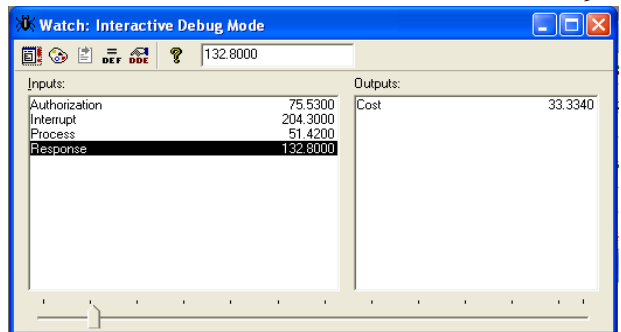

**Figure 14: Input and Output Monitor**

Figure 13 illustrates the process orientation rate that shows how the percentages of process based works in organization maps to fuzzy sets like "too little" or "much". Horizontal vector is started from 0 percentages to 100 percentages and vertical vector is started from 0 to 1 that shows the existence ratio of the point in each set.

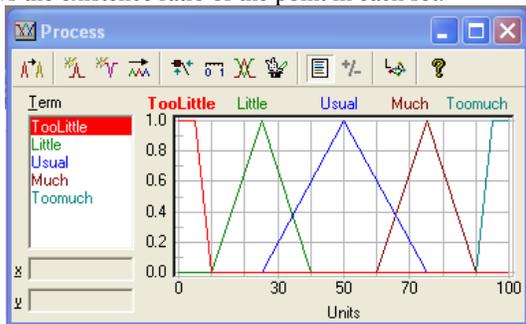

**Figure 13: Process Orientation rate**

Here, is the snap shot of inputting sample data to catch the output by debugging fuzzy black box. The sample data is in a range that had been shown on Table 6. Figure 14 and 15 shows how filling input data interact with black box and how it affects on system.

Figure 14 shows how the inputs will result the output, the output is the cost of support in percentage that starts from 0 to 100 percentage and shows for example if you have such inputs you will consume 33 percentages of support budget.

**Figure 15: Spreadsheet Rule Editor**

As it shows in Figure 15, rules of fuzzy black box on Table 6 have shown here and in this example, rule number one is chosen with the most DoS (about 0.9). Therefore, the Final output is illustrated on Figure 16.

Based on Figure 16, about 33 percentage of budget is predicted for support.

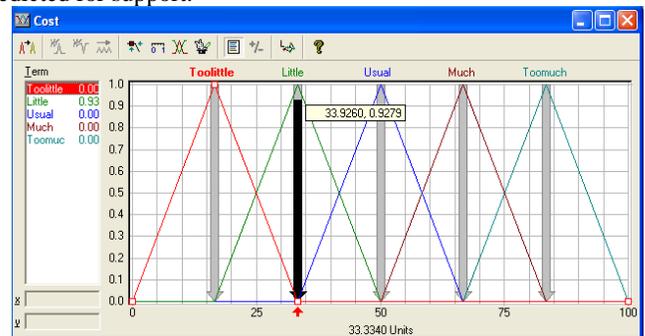

**Figure 16: Cost Prediction**

Figure 16 shows output and how the percentage of the cost of service support can be implied from fuzzy sets like "too little" or "much". Horizontal vector is started from 0 percentages to 100 percentages and vertical vector is started from 0 to 1 that shows the existence ratio of the point in each set.

## 4.4 FRM

As mentioned before in introduction, this paper creates a simple FRM diagram which showed the ITIL processes

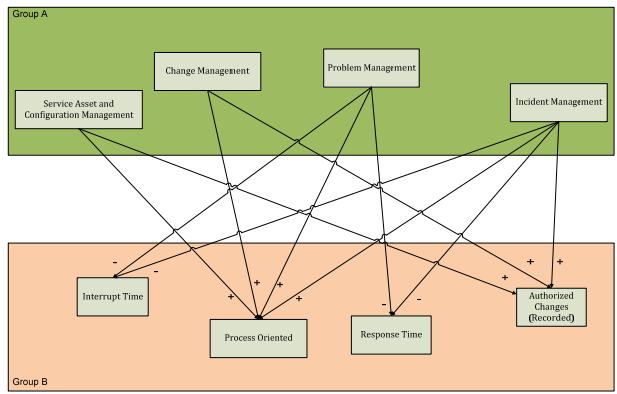

**Figure 17: Group A and Group B**

Figure 17 draws two groups: group 'A' is the ITIL process that implement in service support which were defined in Figure 5 and group 'B' is the reasons or causes (FCM/Fuzzy black box inputs, Figure 6) of implementing of each ITIL processes in group 'A' Therefore, for example, it shows implementation of incident management will result in increasing recording rate and process orientation rate, and it also cause decreasing response time and interrupt time. After drawing FRM diagram it is clear to map the affect off each ITIL process on cost of service support.

➢ Implementation of Incident management will increase the rate of authorized changes.

➢ Implementation of Incident management will increase the rate of process oriented actions.

➢ Implementation of Incident management will decrease the rate of response time to customer.

➢ Implementation of Incident management will decrease the rate of Interrupt time of services.

➢ Implementation of Problem management will increase the rate of process oriented actions.

➢ Implementation of Problem management will decrease the rate of response time to customer.

➢ Implementation of Problem management will decrease the rate of Interrupt time of services.

➢ Implementation of Change management will increase the rate of authorized changes.

➢ Implementation of Change management will increase the rate of process oriented actions.

➢ Implementation of Service Asset and Configuration management will increase the rate of authorized changes.

➢ Implementation of Service Asset and Configuration management will increase the rate of process oriented actions.

## 4.5 Merge FCM with FRM

The main goal of this paper is to find the way to decrease the cost of services (the cost of supporting) by implementing ITIL, so it should define a metrics that affect on costs. The metrics had been defined with FCM that are known "cost of support (final output), response time to customer, Interrupt time of services, the number of process oriented actions, number of recorded events, and number of authorized changes" and then FCM concepts would be normalized to "cost of support (final output), response time to customer, Interrupt time of services, the number of process oriented actions, and number of authorized changes". The number of recorded events was deleted because the number of authorized changes can provide the same meaning of recorded events (it is not important here).

The concepts are:

**First/Main concept:** Cost of service support (most important and final output)

**Metrics group**: "response time to customer, Interrupt time of services, the number of process oriented actions, and number of authorized changes on cost of service"

**ITIL group:** "Incident, Problem, Change and Service Assets and Configuration Management" (ITIL processes)

Here, the first FCM can only help to define the effect of "response time to customer, Interrupt time of services, the number of process oriented actions, and number of authorized changes" on cost of service. Moreover, if you want to find that how ITIL processes can affect directly on the metrics you should define another diagram. It can be FCM, but FRM was preferred because the metric and the ITIL group are two different groups and FRM can provide a better view for their relationship. So, FRM give a view that demonstrates how ITIL processes can affect on secondary metrics and by that it will relate ITIL to output.

**Applicable example:** An organization wants to know how implementing change management can helps to decrease cost of service support; hence, it use FRM model in this paper to find out that change management process could increase the rate of process oriented works; besides, rate of authorized changes. Now, it defines the current status of the rate of process oriented works and rate of authorized changes (for example if all changes done without authorization put 0), then predicts the value of those two metrics after implementing of ITIL Change management and puts the new values in to the FCM black box engine that designs in this paper, thus, it shows the percent of cost support instead of old value. The scenario is shown in figure 18.

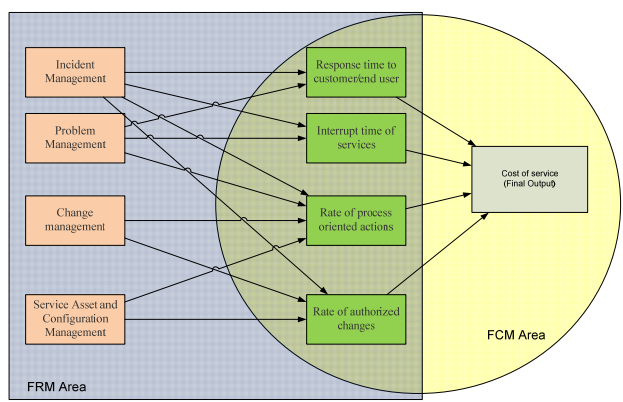

**Figure 18: FRM merge with FCM**

## 5. Conclusion and Future Works

ITIL is a powerful framework that is used to improve the IT activities in organizations by creating process oriented departments to handle such activities. But, implementing a complete version of ITIL may not be comfortable for organizations. Hence, IT managers prefer to implement a light version of ITIL which includes essential process group for their requirements, thus, using an accurate method to define those processes is the main purpose of this paper.

Fuzzy logic has many branches that mostly use in industrial problems but some of branches mostly like Fuzzy Cognitive Map have other usage in different topics. For example, FCM can help to analyze political problems or business problems and so on. Hence, ITIL can be known as problem for FCM to helps it find a method, in order to measure the effects of each process on organization's primarily requirements. This paper is based on ITIL service support approach for an organization, but, there are more approaches that can be analyzed.

As we see, Figure 5 showed processes that are required by Service support, organization's needs and goals had been shown in Figure 6. It was the first FCM diagram that analyzed the relationships between those organization's requirements. After that analyzing the following result is implemented.

Cost of service support is depended on the metrics "Degree of authorization of changes", "Degree of How activities in organization is process oriented", "Metrics related to the response time after each event", "Metrics related to the interrupt time of services after each event".

Those four metrics and designing a fuzzy logic black box help to find the result of inputting items that is termed cost of service support.

Finally, using a FRM diagram showed the relationship between ITIL processes of service support and their effects on FCM's inputs metrics.

There are several methodologies and ideas that define how to implement ITIL and customize it on organizations and all of them do these overall steps:

1. First the ITIL implementer team should find out the current status of organization (AS-IS documents and current service assets that exist)

2. The most important step is to analyze the gaps between current status and what ITIL can do for organization. This step spends the most of time and cost of implementing ITIL, because, it is extremely hard to find out which gaps exist between the current status and ideal status of organization; furthermore, how to fill these gaps with less consuming of time and costs, hence, this paper is written to help implementer team at this step. It is a part of gap analyzes for implementing ITIL that finds out what processes in which way, and how can help organization to catch the goals. For instance, if an organization does not run problem management this paper say if it put problem management in its environment then this process will help to improve the rate of interrupted time between service failures (find out by FRM diagram) and reducing the rate of interrupted time will cause to decrease the cost of service support (find out by FCM and fuzzy black box). This paper just analyzes the gaps for ITIL service support solutions and it is available for most ITIL processes in supporting services as mentions before.

3. After finding the gaps this step helps to find a way from AS-IS to TO-BE base on what found at the previous step.

In a nutshell, this paper can be used as a part of gap analyzes step in implementing ITIL in each organizations.

For the future, the authors will try to works on other parts of ITIL process model approaches and use AI such as Learning automata, PSO algorithms to enhancement the different parameters in ITIL. ITIL helps IT departments to provide a process based work on their organization but how can ITIL help them to do that and the advantages of implementing is the key idea of paper for support solutions. This helps IT managers to define which ITIL processes help them to improve their activities over the organization, so implementation of ITIL in any organization with checking the affect of each disciplines on organization is the result of this paper, but the next steps are to expand this concept for Service Design, Transition and Strategy and show how these disciplines effect on organizations activities.

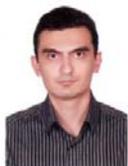 Hamid Zarrazvand Graduated in Bsc from Islamic Azad University of South Tehran branch in computer engineering (Software programming trend) 2004 - 2008, and received his ITIL foundation Certified from Exin 2009 and SEP Sesam Administration Certified from SEP Co 2010.Also, He passed PMI (base on project management body of knowledge) Courses and ITIL Courses at Fanpardaz Institute of Tehran in 2009 and 7-year of work experience in the field of design and implementing Datacenters, Implementing ITIL, COBIT, ISMS for organizations. Moreover, Hamid is specialists in Software programming (ITIL Tools, Office Automations, ERPs and Web applications) in j2se-j2ee-.net framework 3.5,4 and project management.

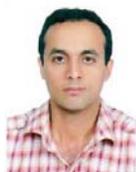 Mohammad Shojafar Received His Bsc in Computer Engineering-Software major Iran University Science and Technology, Tehran, Iran (2001-2006) and Msc in Qazvin Islamic Azad University, Qazvin, Iran (2007-2010). Mohammad is Specialist in Network Programming in Sensor field, Specialist in Distributed and cluster computing (Grid Computing and P2P Computing) , and AI algorithms (PSO, Learning Automata, GA). He is the member of Mechatronics Research Laboratory (MRL) in Iran and Reviewer of International Association of Computer Science and Information Technology (IACSIT) Conference Series and Journal of supercomputing (Springer). He published one Journal in IEEE, one journal in Elsevier, one in JACR, 6 papers in IEEE, and 2 papers in WORLDCOMP Conferences Series held in USA. Mohammad is a Lecturer in Payamnour University in Iran and Analyses and programmer in Exploration Directorate in N .I.O.C Iran now.